\documentclass{article}

\usepackage{amsmath,amssymb}
\usepackage{graphicx}
\usepackage{color}
\usepackage[all]{xy}

\newtheorem{theorem}{Theorem}

\newtheorem{definition}{Definition}

\newtheorem{remark}{Remark}

\newcommand{\ls}[1]
    {\dimen0=\fontdimen6\the\font\lineskip=#1\dimen0
     \advance\lineskip.5\fontdimen5\the\font
     \advance\lineskip-\dimen0
     \lineskiplimit=0.9\lineskip
     \baselineskip=\lineskip
     \advance\baselineskip\dimen0
     \normallineskip\lineskip\normallineskiplimit\lineskiplimit
     \normalbaselineskip\baselineskip
     \ignorespaces}

\begin{document}

\bibliographystyle{abbrv}

\title{Comments on ``A New Method to Compute the 2-Adic Complexity of Binary Sequences"}

\author{Honggang Hu
\smallskip\\
School of Information Science and Technology\\
University of Science and Technology of China\\
Hefei, China, 230027\\
Email. hghu2005@ustc.edu.cn}

\date{}
 \maketitle

\thispagestyle{plain}
\setcounter{page}{1}

\begin{abstract}
We show that there is a very simple approach to determine the 2-adic complexity of periodic binary sequences with ideal two-level autocorrelation. This is the first main result by H. Xiong, L. Qu, and C. Li, IEEE Transactions on Information Theory, vol. 60, no. 4, pp. 2399-2406, Apr. 2014,  and the main result by T. Tian and W. Qi, IEEE Transactions on Information Theory, vol. 56, no. 1, pp. 450-454, Jan. 2010.
\end{abstract}

\ls{1.5}
\section{A Very Simple Approach}

Let $S=\{s_i\}_{i=0}^{+\infty}$ be a periodic binary sequence with period $N$, and $S(x)=\sum_{i=0}^{N-1}s_ix^i\in \mathbb{Z}[x]$. Let us write
$$\frac{S(2)}{2^N-1}=\frac{\sum_{i=0}^{N-1}{{s_i}{2^i}}}{2^N-1}=\frac{p}{q}$$
with $0\leq p\leq q$, and $\gcd(p, q)=1$.

\begin{definition}[\cite{KG97}]
With the notations as above, the 2-adic complexity $\Phi(S)$
of $S$ is the real number $\log_2{q}$.
\end{definition}

\begin{remark}
If $\gcd(S(2), 2^N-1)=1$, then the 2-adic complexity $\Phi(S)$
of $S$ achieves the maximum value $\log_2 (2^N-1)$.
\end{remark}

For any $0\leq\tau<N$, the autocorrelation of $S$ at shift $\tau$ is defined by
$$C_S(\tau)=\sum_{i=0}^{N-1}(-1)^{s_{i+\tau}+s_i}.$$
If $C_S(\tau)=-1$ for any $0<\tau<N$, we call $S$ an ideal two-level autocorrelation sequence \cite{GG2005}. There are three cases of $N$ such that there exists an ideal two-level autocorrelation sequence of period $N$: 1) $N=2^n-1$; 2) $N=p$, where $p$ is a prime number with $p\equiv3\ (\mbox{mod }4)$; 3) $N=p(p+2)$, where both $p$ and $p+2$ are prime numbers \cite{GG2005}.

Let $P(x)=\sum_{i=0}^{N-1}(-1)^{s_i}x^i\in \mathbb{Z}[x]$. If $S$ is an ideal two-level autocorrelation sequence, then we have
\begin{eqnarray*}
P(x)P(x^{-1})&=&\left(\sum_{i=0}^{N-1}(-1)^{s_i}x^i\right)\left(\sum_{j=0}^{N-1}(-1)^{s_j}x^{-j}\right)\ (\mbox{mod }x^N-1)\\
&=&\sum_{i=0}^{N-1}\sum_{j=0}^{N-1}(-1)^{s_i+s_j}x^{i-j}\ (\mbox{mod }x^N-1)\\
&\equiv&N+\sum_{\tau=1}^{N-1}\sum_{j=0}^{N-1}(-1)^{s_{j+\tau}+s_j}x^{\tau}\ (\mbox{mod }x^N-1)\\
&\equiv& N-x-x^2-...-x^{N-1}\ (\mbox{mod }x^N-1).
\end{eqnarray*}

As a consequence, we have
\begin{eqnarray*}
P(2)P(2^{-1})&\equiv&N-2-2^2-...-2^{N-1}\ (\mbox{mod }2^N-1)\\
&\equiv&N+1\ (\mbox{mod }2^N-1).
\end{eqnarray*}
Note that $P(2)=\sum_{i=0}^{N-1}(-1)^{s_i}2^i=\sum_{i=0}^{N-1}(1-2s_i)2^i=2^N-1-2\cdot S(2)$. Hence, we obtain the following interesting theorem.

\begin{theorem}
With the notations as above, we have
$$S(2)P(2^{-1})\equiv-\frac{N+1}{2}\ (\mbox{mod }2^N-1).$$
\end{theorem}

By a simple argument, we can show that $\gcd(N+1, 2^N-1)=1$ as did in \cite{XQL14}. It follows that $\gcd(S(2), 2^N-1)=1$ which means that the 2-adic complexity of such sequences is maximum.

\section{Conclusion}

Using the property of ideal two-level autocorrelation carefully, we find a very simple way to show that the 2-adic complexity of ideal two-level autocorrelation sequences is maximum. This is the main result in \cite{TQ10}, and the first main result in \cite{XQL14}. For the case of symmetric 2-adic complexity \cite{HF08}, the same result also holds as pointed out in \cite{XQL14}.



\begin{thebibliography}{100}

\bibitem{GG2005}
S. Golomb and G. Gong, {\em Signal Designs With Good Correlation:
For Wireless Communications, Cryptography and Radar Applications}.
Cambridge, U.K.: Cambridge University Press, 2005.

\bibitem{HF08}
H. Hu and D. Feng, ``On the 2-adic complexity and the $k$-error 2-adic complexity of periodic binary sequences," {\em IEEE Trans. on Inform. Theory}, vol. 54, no. 2, pp. 874-883, Feb. 2008.

\bibitem{KG97}
A. Klapper and M. Goresky, ``Feedback shift registers, 2-adic span, and
combiners with memory," {\em J. Crypt.}, vol. 10, pp. 111-147, 1997.

\bibitem{TQ10}
T. Tian and W. Qi, ``2-adic complexity of binary $m$-sequences," {\em IEEE Trans. on Inform. Theory}, vol. 56, no. 1, pp. 450-454, Jan. 2010.

\bibitem{XQL14}
H. Xiong, L. Qu, and C. Li, ``A new method to compute the 2-adic complexity of binary sequences," {\em IEEE Trans. Inf. Theory}, vol. 60, no. 4, pp. 2399-2406, Apr. 2014.

\end{thebibliography}
\end{document}